\documentclass[a4paper, 9pt, twocolumn]{article}
\usepackage{geometry}                
\usepackage{graphicx}
\usepackage{amssymb}
\usepackage{amsthm}
\usepackage{epstopdf}
\usepackage{mathtools}
\usepackage[shortlabels]{enumitem}
\usepackage{dsfont}
\usepackage{cuted}
\usepackage{sidecap}
\usepackage[usenames,dvipsnames]{color}
\usepackage[thinspace,mediumqspace,Grey,squaren]{SIunits}
\usepackage{bbm}
\usepackage[T1]{fontenc}
\usepackage[utf8]{inputenc}
\usepackage{authblk}
\usepackage[font=small,labelfont=bf ,textfont={small, sf}]{caption}
\usepackage{titlesec}
\usepackage{url}
\usepackage{hyperref}
\titleformat{\subsection}[runin]{\normalfont\bfseries}{\thesubsection.}{3pt}{}

\def\ExtrinsicSpace{\mathcal{Z}}

\def\NumReactions{N}
\def\NumSpecies{M}
\def\Propensity{h}
\def\MarginalPropensity{\lambda}

\def\JointMarkovChain{Y}
\def\ExtrinsicMarkovChain{Z}
\def\IntrinsicMarkovChain{X}
\def\JointState{Y}

\def\ExtrinsicState{Z}
\def\extrinsicState{z}

\def\ExtrinsicStateInt{\mathbf{\ExtrinsicState}}
\def\IntrinsicState{X}
\def\intrinsicState{x}
\def\intrinsicStateInt{\mathbf{\intrinsicState}}
\def\IntrinsicStateInt{\mathbf{\IntrinsicState}}
\def\KolmogorovOperator{\mathcal{A}}
\def\DifferentialOperator{\mathcal{D}}

\def\moment{M}
\def\centralMoment{S}
\newcommand{\momentOrder}[1]{\moment_{#1}}
\newcommand{\centralMomentOrder}[1]{\centralMoment_{#1}}

\def\State{X}

\def\StateInt{\mathbf{\State}}
\def\state{x}
\def\stateInt{\mathbf{\state}}

\def\kparameters{c}
\def\kparameter{c}

\def\Real{\mathbb{R}}

\newcommand{\Expect}[1]{ \mathbb{ E } \left[ #1 \right]}

\newcommand{\Var}[1]{ \mathrm{Var} \left[ #1 \right]}

\begin{document}

\title{Uncoupled analysis of stochastic reaction networks in fluctuating environments}

\author[1]{Christoph Zechner}
\author[1,2]{Heinz Koeppl}
\affil[1]{Automatic Control Laboratory, ETH Zurich, Switzerland}
\affil[2]{Technische Universit\"at Darmstadt, Darmstadt, Germany}
\renewcommand\Authands{ and }

\date{}

\maketitle

\newpage{}

{\noindent \sf \fontsize{9}{0} \selectfont \textbf{
The dynamics of stochastic reaction networks within cells are inevitably modulated by factors considered extrinsic to the network such as for instance the fluctuations in ribsome copy numbers for a gene regulatory network. While several recent studies demonstrate the importance of accounting for such extrinsic components, the resulting models are typically hard to analyze. 
In this work we develop a general mathematical framework that allows to uncouple the network from its dynamic environment by incorporating only the environment's effect onto the network into a new model. More technically, we show how such fluctuating extrinsic components (e.g., chemical species) can be marginalized in order to obtain this decoupled model. We derive its corresponding process- and master equations and show how stochastic simulations can be performed. Using several case studies, we demonstrate the significance of the approach. For instance, we exemplarily formulate and solve a marginal master equation describing the protein translation and degradation in a fluctuating environment.
}}

\section*{Introduction}
Biochemical systems involving low-copy molecules demand for mathematical models that account for the intrinsic stochasticity \cite{Mcadams1997}. In recent years, however, realization has grown that intrinsic noise alone cannot account for the observed substantial phenotypic variability among isogenic cells. That is, fluctuations in the intracellular environment, commonly termed extrinsic noise, represent an additional source of variability \cite{Elowitz2002, Colman-Lerner2005, oshea}. 

Several recent studies focus on separating intrinsic and extrinsic fluctuations through dual-reporter measurements \cite{Hilfinger2011, Swain2002}. Other approaches model extrinsic noise through certain parameters (i.e., the translation rate) of a kinetic model which is calibrated subsequently using flow-cytometry \cite{Ruess2013, Zechner2012} or time-lapse microscopy data \cite{Zechner2014, Finkenstadt2013}. All of those approaches have in common that they consider the biochemical process under study -- i.e., the expression of a gene -- as a small subpart that is embedded into a larger dynamical system. Accordingly, they rely on augmenting the original kinetic model by certain environmental components which are assumed to be fixed but random \cite{Hasenauer2011, Finkenstadt2013, Zechner2012, Zechner2014} or fluctuating over time \cite{Hilfinger2011, Shahrezaei2008}. In fact, such models agree very well with the variability that is observed experimentally, but on their downside, suffer from the increased dimensionality - somehow defeating the original purpose of tractable dedicated models. 

A natural question arising in that context is whether we can find a proper dynamical description of just the system of interest as if it was still embedded into its stochastically modulating environment. 
In other words, we aim to find a ``self-contained'' stochastic model that summarizes all system behaviors attainable under all possible realizations of the extrinsic fluctuations. Such models could then be used to perform an uncoupled analysis of a reaction network subject to extrinsic noise. The mathematical correct answer that we provide in this work is the marginalization of the system dynamics with respect to those extrinsic fluctuations. Interestingly it turns out that the resulting model exploits its own stochasticity to emulate the effect of extrinsic noise, leading to a self-exciting process. A simple instance of such self-excitation is the Polya urn scheme\footnote{At each draw from the Polya urn with balls of two colors the drawn ball and a fixed number of new balls of the same color as the draw are placed in the urn.}, which is known to be equivalent to Bernoulli trials marginalized over random (and here correspondingly extrinsic) success rates \cite{Johnson1977}. Intuitively, due to its self-excitation, the number of draws of the same color for a Polya urn over repeated draws displays a much richer and dispersed dynamics than the number of sucessful draws in a Bernoulli trial with fixed success rate.   

For the purpose of inference we recently proposed a first attempt of such marginalization for the special case of fixed but random environmental conditions \cite{Zechner2014}. In this work we develop a general mathematical framework from which the uncoupled dynamics can be constructed in a principled manner, regardless whether the environment is constant or dynamically changing.

\section*{Results}

\subsection*{Mathematical modeling}
We describe the time-evolution of a stochastic reaction network by a continous-time Markov chain (CTMC) $\IntrinsicMarkovChain$ with $\NumSpecies$ chemical species and $\NumReactions$ reaction channels.
The system state at time $t$ is denoted $\IntrinsicState(t)$ and we write its random\footnote{Throughout we will follow the usual convention to refer to upper-case and lower-case versions of a symbol as a random variable and its realization, respectively.} path on time intervals $[0, a]$ as $\IntrinsicStateInt_{a}$. Furthermore, we assume that $\IntrinsicMarkovChain$ depends on another multivariate Markov process $\ExtrinsicMarkovChain$ through its hazard functions in the form 
\begin{equation}
	\Propensity_{i}(\intrinsicState, \extrinsicState) = \kparameter_{i}(\extrinsicState) g_{i}(\intrinsicState),
	\label{eq:ModulatedPropensity}
\end{equation} 
with $\kparameter_{i}$ some positive function and $g_{i}$ a polynomial determined by the law of mass-action, for instance. For reactions independent of $\ExtrinsicMarkovChain$, we thus have $c_{i}(\extrinsicState)\equiv c_{i}$. Typically, $\ExtrinsicMarkovChain$ is another jump or diffusion process corresponding to a set of modulating \textit{environmental} species or conditions that are considered extrinsic to the system of interest, whereas the species in $\IntrinsicMarkovChain$ represent the actual system of interest. For example, $\ExtrinsicMarkovChain$ could be the fluctating ribosome copy numbers affecting the kinetics of a gene regulatory network represented by $\IntrinsicMarkovChain$. Although a more general treatment is possible, we assume a feed-forward structure between $\ExtrinsicMarkovChain$ and $\IntrinsicMarkovChain$, which means that  $\ExtrinsicMarkovChain$ modulates $\IntrinsicMarkovChain$ but not vice-versa. Consequently, the dynamics of the joint system $\JointState(t) = \left(\ExtrinsicState(t), \IntrinsicState(t)\right)$ can be described by a marginal Markov process $\ExtrinsicState$ together with a conditional Markov chain $\IntrinsicState \mid \ExtrinsicState$.

\subsection*{Uncoupled dynamics}
Mathematical descriptions of the joint system $\JointState(t)$ are readily obtained using available techniques for modeling Markovian dynamics \cite{Hilfinger2011, Koeppl2012, Shahrezaei2008}. For complexity reasons, however, we aim for models that can properly describe \textit{only} the interesting components $\State(t)$. In order to see that marginalization over $\ExtrinsicMarkovChain$ yields the desired model, let us first consider two dependent random variables $A$ and $B$ described by a joint probability distribution $p(a, b) = p(a\mid b) p(b)$. If we are interested in analyzing $A$ under all possible values of $b$, we need to average the probability at $A=a$ over all possible values of $b$, i.e., $$p(a) = \int p(a, b) \mathrm{d}b = \Expect{p(a \mid B)}.$$ 
	Note that as a consequence of averaging probabilities, any value $a$ possible (i.e. has non-zero measure) under the joint $p(a,b)$ is possible under the marginal $p(a)$, while this does not necessarily apply to $p(a \mid b)$ for any choice of $b$.  
 
In case of the coupled processes $\ExtrinsicState(t)$ and $\IntrinsicState(t)$, we analogously \textit{marginalize} the joint Markov chain $\JointMarkovChain$ with respect to the environmental process $\ExtrinsicState$. While such a marginalization involves several difficulties, the idea remains the same: we try to construct an uncoupled process $\IntrinsicMarkovChain$ which directly admits the marginal path distribution $p(\stateInt_{t}) = \Expect{p(\stateInt_{t}\mid \ExtrinsicStateInt_{t})}$, bypassing the intractable averaging over all possible extrinsic histories. As a result, we obtain a jump process which - in contrast to the conditional process $\IntrinsicMarkovChain \mid \ExtrinsicMarkovChain$ - no longer depends on the environmental species in $\ExtrinsicMarkovChain$. We remark that a straightforward marginalization of the joint master equation of $\ExtrinsicMarkovChain$ and $\IntrinsicMarkovChain$ generally leads to intractable propensities \cite{Arkin2003, Hilfinger2011}.
Based on the innovation theorem \cite{AalenBook} we demonstrate in section S.1 in the SI Appendix that the hazard functions of the uncoupled process can be generally written as
\begin{equation}
	\MarginalPropensity_{i}(\IntrinsicStateInt_{t}) =  \Expect{\kparameter_{i}\left(\ExtrinsicState(t) \right) \mid \IntrinsicStateInt_{t}} g_{i}(\IntrinsicState(t)),
	\label{eq:GeneralMarginalPropensity}
\end{equation}
where the expectation is taken with respect to the conditional distribution $\pi\left(\extrinsicState, t \mid \intrinsicStateInt_{t}\right)$. The latter describes the conditional probability of the environmental process $\ExtrinsicState(t)$ given the entire history of process $\IntrinsicMarkovChain$ until time $t$\footnote{More precisely we would need to say that $\IntrinsicStateInt_{t}$ is a filtration of $\IntrinsicMarkovChain$.}. Using the expected value of that distribution, the feed-forward influence of $\ExtrinsicMarkovChain$ on the hazard functions of $\IntrinsicMarkovChain$ can be replaced by a deterministic function of $\IntrinsicMarkovChain$, which no longer depends on the actual state of $\ExtrinsicMarkovChain$. Instead, the marginal process $\IntrinsicMarkovChain$ becomes \textit{self-exciting}, meaning that it exerts a feedback on itself. Note that the uncoupled process $\IntrinsicMarkovChain$ is no longer Markovian, since the conditional expectation - and hence the hazard functions - depend on the full process history $\IntrinsicStateInt_{t}$. A schematic illustration of that uncoupling is given in Fig.1.

\begin{figure}
	\centering
	\includegraphics{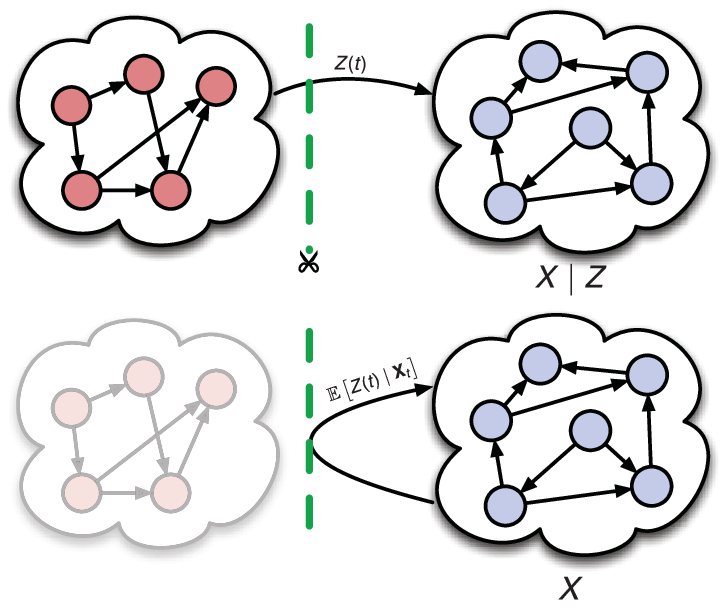}
	\caption{Uncoupled stochastic dynamics. The environmental process $\ExtrinsicMarkovChain$ modulates the dynamics of the process under study $\IntrinsicMarkovChain$, e.g., through one of its hazard functions. Marginalization with respect to $\ExtrinsicMarkovChain$ yields the uncoupled dynamics of $\IntrinsicMarkovChain$, whereas the original dependency on the environment $\ExtrinsicMarkovChain$ is replaced by its optimal estimator given the history of $\IntrinsicMarkovChain$. Consequently, the marginal process $\IntrinsicMarkovChain$ is self-exciting, i.e., it exerts a feedback on itself.}
		\label{fig:Scheme}
\end{figure}

\subsection*{Solving the accompanying filtering problem}
Although the construction of the uncoupled dynamics is general, any practical implementation thereof will depend on an explicit computation of the conditional expectation in Eq.~\ref{eq:GeneralMarginalPropensity}. This expectation estimates the environmental state $\ExtrinsicState(t)$ given the full history of the uncoupled process $\StateInt_{t}$ and therefore, can be understood as the solution to a \textit{stochastic filtering} problem \cite{crisanbook}.
Filtering techniques deal with the problem of optimally reconstructing a hidden stochastic process at time $t$ from noisy observations of that process up to time $t$. 
In the situation considered here, the hidden process corresponds to the environment $\ExtrinsicState(t)$, which gets reconstructed from the ``observed'' history $\IntrinsicStateInt_{t}$ through the conditional mean in Eq.~\ref{eq:GeneralMarginalPropensity}.

We assume that the environment $\ExtrinsicState(t)$ admits a probability distribution $p(\extrinsicState, t)$ described by a Kolmogorov-forward equation of the form
\begin{equation} 
	\frac{\partial}{\partial t}p(\extrinsicState, t) = \KolmogorovOperator p(\extrinsicState, t),
	\label{eq:KolmogorovForward}
\end{equation}
where $\KolmogorovOperator$ represents the temporal change of $p(\extrinsicState, t)$, i.e., is the infinitesimal generator of $\ExtrinsicMarkovChain$. For instance, if $\ExtrinsicMarkovChain$ is a diffusion process, $\KolmogorovOperator$ corresponds to the Fokker-Planck operator, while in case of a CTMC, $\KolmogorovOperator$ is given by the difference operator of the chemical master equation (CME). In terms of filtering, Eq. \ref{eq:KolmogorovForward} corresponds to the process model of $\ExtrinsicMarkovChain$. 
Furthermore, we know that at a given time $t$, the solution of $\IntrinsicMarkovChain$ can be written as a sum of independent but time-transformed Poisson processes \cite{Kurtz2011}, each of them corresponding to a particular reaction channel. Consequently, the observation model is given by a set of Poisson counting observations with the hazard functions given in Eq.\ref{eq:ModulatedPropensity}. This is closely related to Markov-modulated Poisson processes \cite{Snyder} and their corresponding optimal filtering \cite{Elliott2005}. 

While a more general treatment is provided in the SI Appendix, we assume in the following that a one-dimensional process $\ExtrinsicMarkovChain$ is modulating $\IntrinsicMarkovChain$ through its $k$-th reaction of order zero. We further restrict ourselves to the case where $\kparameter_{k}$ is a linear function of $\extrinsicState$, i.e., $\kparameter_{k}(\extrinsicState) = \kparameter_{k} \extrinsicState$. Under those assumptions, it can be shown that the conditional process $\ExtrinsicMarkovChain(t) \mid \IntrinsicStateInt_{t}$ follows a filtering distribution $\pi(\extrinsicState, t \mid \intrinsicStateInt_{t}) \equiv \xi(t) \tilde{\pi}(\extrinsicState, t)$ with
\begin{equation}
	\begin{split}
	\mathrm{d} \tilde{\pi}(\extrinsicState, t)  = \left[\KolmogorovOperator \tilde{\pi}(\extrinsicState, t)- \kparameter_{k}  \extrinsicState \tilde{\pi}(\extrinsicState, t) \right] \mathrm{d}t  +  \left[ \extrinsicState - 1\right]  \tilde{\pi}(\extrinsicState, t)  \mathrm{d} R_{k}(t),
	\end{split}
	\label{eq:PosteriorProbability}
\end{equation}
with $\xi(t)$ a time-dependent normalizing factor independent of $\extrinsicState$ and $R_{k}(t)$ the number of reactions of type $k$ up to time $t$ in $\IntrinsicStateInt_t$. Thus, Eq.\ref{eq:PosteriorProbability} describes a scaled version of the normalized filtering distribution. The latter shows an implicit dependency on its own mean (see Methods and section S.2 in the SI Appendix) and is therefore complicated to handle numerically. In contrast, once we have numerically solved for $\tilde{\pi}$, it can be easily rescaled such that it integrates (or sums up) to one for all $t$. Note that Eq. \ref{eq:PosteriorProbability} is a stochastic partial differential equation (SPDE) in case $\ExtrinsicMarkovChain$ describes a diffusion process or a stochastic difference-differential equation (SDDE) if $\ExtrinsicMarkovChain$ is a CTMC. 
In the latter case, the solution of Eq.\ref{eq:PosteriorProbability} can be compactly written as
\begin{equation}
	\tilde{\Pi}(t) = e^{(Q - c_{k} \Lambda ) t} \Lambda^{R_{k}(t)} \Pi_{0}, 
	\label{eq:FilteringDistributionClosedForm}
\end{equation}
with $\tilde{\Pi}(t) = (\tilde{\pi}(0, t), \ldots, \tilde{\pi}(L-1, t))^{T}$, $L$ the number of reachable states of $\ExtrinsicMarkovChain$, $Q \in \Real^{L\times L}$ the generator matrix of $\ExtrinsicState$, $\Lambda = \mathrm{diag}(0, \ldots, L-1)$ and $\Pi_{0} \in \Real^{L}$ the initial distribution over $\ExtrinsicState$.

In order to evaluate Eq. \ref{eq:GeneralMarginalPropensity}, we only require the mean (i.e., the first moment) of the filtering distribution, i.e., $\momentOrder{1}(t) = \Expect{\ExtrinsicState(t) \mid \IntrinsicStateInt_{t}}$. In general, however, the mean also depends on the second-order moment, which in turn depends on the third-order moment and so forth. We show in the Methods section that the (non-central) filtering moment dynamics up to order $i$ can be generally written as
\begin{equation}
	\begin{split}
		&\mathrm{d} \momentOrder{1}(t) = \left[\DifferentialOperator_{1}(t) - \kparameter_{k} (\momentOrder{2}(t) - \momentOrder{1}(t)\momentOrder{1}(t)) \right]\mathrm{d}t \\
		&\quad\quad\quad\quad + \frac{\momentOrder{2}(t) - \momentOrder{1}(t)\momentOrder{1}(t)}{\momentOrder{1}(t)} \mathrm{d}R_{k}(t) \\
		& \quad  \quad \quad \quad \vdots \\
		&\mathrm{d} \momentOrder{i}(t)= \left[\DifferentialOperator_{i}(t) - \kparameter_{k} (\momentOrder{i+1}(t) - \momentOrder{1}(t) \momentOrder{i}(t)) \right]\mathrm{d}t \\
		&\quad\quad\quad\quad + \frac{\momentOrder{i+1}(t) - \momentOrder{1}(t)\momentOrder{i}(t)}{\momentOrder{1}(t)}\mathrm{d}R_{k}(t),
	\end{split}
	\label{eq:momentSystem}
\end{equation}
where $\DifferentialOperator_{j}(t)$ refers to the prior dynamics of the $j$-th moment. Although Eq.\ref{eq:momentSystem} is generally infinite-dimensional, there are several relevant scenarios, for which the moment dynamics are \textit{closed}, i.e., only depend on higher-order moments up to a certain order. This is for instance the case, if $\ExtrinsicState(t)$ is a Cox-Ingersoll-Ross process or any finite state Markov chain. On the other hand, if the moment dynamics are infinite-dimensional, suitable assumptions on the filtering distribution $\pi$ can be imposed to yield a closed moment-dynamics (see S.3 in the SI Appendix). An important closure is found by analyzing Eq.~\ref{eq:FilteringDistributionClosedForm}: especially for large $\kparameter_{k}$, the conditional distribution of $\ExtrinsicMarkovChain$ is predominantly driven by the term $e^{-\kparameter_{k} \Lambda t}\Lambda^{R_{k}(t)}$, suggesting that it can be well approximated by a Gamma-distribution. We note that the Gamma-distribution is fully characterized by two parameters -- or equivalently -- its first two moments $\momentOrder{1}(t)$ and $\momentOrder{2}(t)$. As a consequence, we may express the third order moment as a function of the first two moments, i.e., $\momentOrder{3}(t) = -\momentOrder{1}(t) \momentOrder{2}(t) + 2 \momentOrder{2}^2(t)/\momentOrder{1}(t)$, such that the second conditional moment closes as
\begin{align}
	&\mathrm{d} \momentOrder{2}(t) = \left[\DifferentialOperator_{2}(t) - 2 \kparameter_{k} \frac{\momentOrder{2}(t)}{\momentOrder{1}(t)}\left(\momentOrder{2}(t) - \momentOrder{1}^{2}(t)\right) \right]\mathrm{d}t \\
		 &\quad\quad\quad\quad+ 2\left[\frac{\momentOrder{2}^{2}(t)}{\momentOrder{1}^2(t)} - \momentOrder{2}(t)\right] \mathrm{d}R_{k}(t).
		 \nonumber
\end{align}
Further discussion on this closure is provided in section S.3 in the SI Appendix. 

\subsection*{Stochastic simulation} 

Although the uncoupled dynamics of $\IntrinsicMarkovChain$ are non-Markovian, the Markov property can be enforced by virtually extending the state space by the filtering mean of Eq.~\ref{eq:momentSystem}, which summarize the history of $\IntrinsicMarkovChain$. As a result, one can simulate sample paths of the uncoupled process using standard methods that can account for the explicit time-dependency of the hazard functions \cite{Anderson2007}. In general, such algorithms rely on the generation of random waiting times for each of the reaction channels. All reactions that are independent of $\ExtrinsicState(t)$ will retain their exponentially distributed waiting times. In contrast, the time $\tau_{k}$ that passes until a reaction of type $k$ happens is distributed according to
\begin{equation}
	P_{k}(\tau_{k} < s \mid \StateInt_{t}) = 1 - e^{-\kparameter_{k} \int_{0}^{s}\momentOrder{1}(t+T) \mathrm{d}T}.
	\label{eq:WaitingTimeGeneral}
\end{equation}
We note that as long as no reaction of type $k$ happens, $\mathrm{d}R_{k}(t)$ is zero and hence, $\momentOrder{1}(t)$ is found by solving a set ordinary differential equations (ODEs). Since that solution is not generally known in closed form, we cannot directly sample from Eq. \ref{eq:WaitingTimeGeneral}. However, several efficient solutions to that problem have been developed in the context of inhomogeneous Poisson processes, e.g., such as the method of \textit{thinning} \cite{Lewis1979} (see Methods). Once a reaction has fired, the filtering moments need to be updated by the terms multiplying the firing process $\mathrm{d}R_{k}(t)$ in Eq.~\ref{eq:momentSystem} (i.e., they exhibit a discontinuity). 

Evidently, simulation from Eq. \ref{eq:WaitingTimeGeneral} comes at higher cost than simulating from an exponential distribution (e.g., such as performed in standard SSA algorithms), since in general, it relies on the numerical integration of an ODE. However, reactions associated with the environmental part no longer need to be simulated, which yields a significant reduction in computational effort as soon as the environmental network is large and expensive to simulate due to high propensity reactions, for instance.

\subsection*{Fluctuations on different timescales} The impact of environmental fluctuations on a dynamical system of interest is as diverse as the timescale on which they operate. For instance, extrinsic noise in the context of gene expression might be slowly varying (e.g., correlates well with the cell-cycle \cite{Rosenfeld2005, Volfson2006}), while fluctuations in transcription factor abundance might be significantly faster than the expression kinetics downstream. From a technical point of view, timescales range from constant environmental conditions that are random but fixed \cite{Zechner2013} to regimes where the fluctuations are very fast, such that quasi-steady-state (QSS) assumptions become applicable \cite{Arkin2003}. A QSS-based approach for simulating a system $X$ in the presence of extrinsic noise $Z$ corresponds to simulating the conditional CTMC $ X \mid Z$, where $Z$ is replaced by the mean of $Z$. 
The simulation of the joint system $(X,Z)$ become prohibitive if extrinsic fluctuations are fast, while with Eq.~\ref{eq:momentSystem} the complexity of the marginal process simulation is invariant with respect to the time-scale of the environment. Alternatively, one may try to replace a fluctuating environment $Z$ through a random but fixed enviroment of same variance but this leads to an overestimation of the process variance in $X$ \cite{Hilfinger2011}, as discussed in a later section.  
To investigate the two above simplifying assumptions and compare them to the exact solution obtained via SSA and via the marignal process, we performed a simulation study on a linear three-stage birth-death model given in Fig.2a, where only species \textit{C} is considered of interest in this case. Accordingly, the uncoupled dynamics of \textit{C} are obtained by integrating the dynamics over the \textit{A} and \textit{B}. The results are shown in Fig.2b and Fig.2c.

\begin{figure}
	\centering
	\includegraphics[width=0.9\columnwidth]{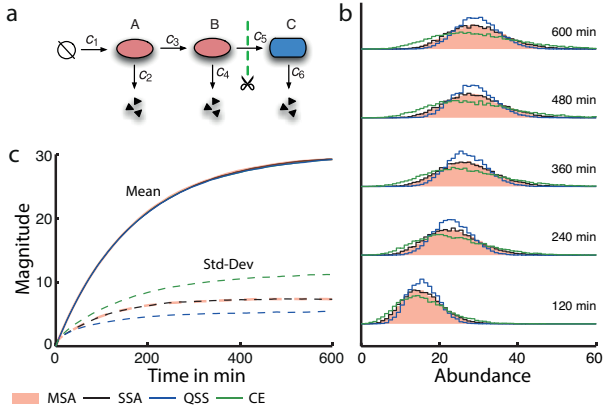}
	\caption{Marginal simulation algorithm. (a) Simple three-stage model. Species A, B and C are modeled as coupled linear birth-death processes, where the coupling is realized by linearly modulating the birth rates of B and C (rate constants $c_1=0.003, c_2=0.001, c_3=0.05, c_4=5e-4, c_5=1e-5, c_6=1e-4$). The uncoupled marginal dynamics of C are obtained by integrating over fluctuations of species A and B (7000 sample paths were used). (b, c) Evaluation of the marginal simulation algorithm. Simulations based on the QSS-approximation neglect a significant portion of variability as opposed to assuming a constant environment (CE) in which case the variability is overestimated. In contrast, the uncoupled dynamics correctly predict the fluctuations on the protein level, while yielding a reduction in computational effort when compared to standard SSA (20min simulation time instead of 46min); correspondingly higher speedup can be achieved for a larger time-scale separation of processes (A,B) versus C.}
		\label{fig:MarginalSimulation}
\end{figure}

\subsection*{Propagation of environmental fluctuations and the effective noise} 

Several recent studies \cite{Bowsher2013, Hilfinger2011, Swain2002, oshea} are centered around the separation of different noise contributions in biochemical networks. Typically, the law of total variance is employed to decompose the fluctuations of $\State(t)$ into parts that are intrinsic to $\State$ and parts that come from $\ExtrinsicState$ (i.e., are extrinsic to $\State$). Here we found that performing such an analysis on $\ExtrinsicState$ instead of $\State$ -- in conjunction with our decoupling approach -- provides a novel way to study how stochasticity is propagated through biochemical networks. Using the law of total variance, we can decompose the total (or unconditional) variance of $\ExtrinsicState(t)$ as
\begin{equation}
	\Var{\ExtrinsicState(t)} = \Expect{\Var{Z(t) \mid \StateInt_{t}}} + \Var{\Expect{Z(t) \mid \StateInt_{t}}}.
	\label{eq:VarianceDecomposition}
\end{equation}
The two terms on the r.h.s. can be interpreted as follows. Assume we can observe $\ExtrinsicState$ only through $\State$. Since $\State$ is intrinsically stochastic, a part of the variability of $\ExtrinsicState$ is not carried over to $\State$. In Eq. \ref{eq:VarianceDecomposition}, this part (i.e., the \textit{suppressed noise}) corresponds to the first term on the r.h.s. since it quantifies the uncertainty about $\ExtrinsicState(t)$ that remains after observing $\StateInt_{t}$. The second term determines how accurate $\ExtrinsicState$ can be reconstructed from trajectories of $\StateInt$. Alternatively, it can be understood as the amount of noise in $\ExtrinsicState$ that effectively impacts $\State$ (i.e., the \textit{effective noise}). For instance, the environmental process could be characterized by a large variance, but still have only marginal impact on $\State(t)$ -- depending on the timescale of $\ExtrinsicState$ and $\IntrinsicState$. 

In order to quantify those terms, we note that the conditional variance within in the first term coincides with the second-order central moment of the filtering distribution from Eq.\ref{eq:PosteriorProbability}. This further implies that it can be computed  ``on-the-fly'' when simulating $\State(t)$ using the marginal simulation algorithm which allows an efficient estimation of its expectation. However, in some biologically relevant cases, the effective noise can be determined even analytically, which we demonstrate in the following. 

We derive in section S.4 in the SI Appendix that the expected central moments are generally given by
\begin{equation}
	\begin{split}
	\frac{\mathrm{d}}{\mathrm{d}t} \Expect{\momentOrder{1}(t)} &= \Expect{\DifferentialOperator_{1}(t)} \\
	\frac{\mathrm{d}}{\mathrm{d}t} \Expect{\centralMomentOrder{2}(t)} &= \Expect{\tilde{\DifferentialOperator}_{2}(t)} - c_{k} \Expect{\frac{\centralMomentOrder{2}^{2}(t)}{\momentOrder{1}(t)}}.
	\label{eq:ExpectedMoments}
\end{split}
\end{equation}
The mean in Eq.\ref{eq:ExpectedMoments} is just the unconditional mean of $\ExtrinsicState(t)$, while the derivative of the expected variance shows an additional negative term, causing it to be smaller than the unconditional variance. 
Let us for instance consider the case where $\ExtrinsicState(t)$ follows a Cox-Ingersoll-Ross (CIR) process governed by the SDE 
\begin{equation}
	\mathrm{d}\ExtrinsicState(t) = \theta ( \mu - \ExtrinsicState(t) ) \mathrm{d}t + \sigma_{\ExtrinsicState} \sqrt{\ExtrinsicState(t)} \mathrm{d}W(t),
\end{equation}
with $\theta$, $\mu$ and $\sigma_{\ExtrinsicState}$ as real process parameters and $W(t)$ as a standard Wiener process. 
Note that in this case, Eq.~\ref{eq:ExpectedMoments} reduces to an autonomous ODE, which for large $t$ yields the relative effective noise $\eta = \Var{\Expect{\ExtrinsicState(t)\mid \StateInt_{t}}}/ \Var{\ExtrinsicState(t)}$ at stationarity, i.e., 
\begin{equation}
	\eta = 1 + 2 \frac{v^{2}}{\kparameter_{k}} \left(1 - \sqrt{\frac{\kparameter_{k}}{v^{2}}+1} \right),
	\label{eq:EffectiveNoise}
\end{equation}
where $v=\theta / \sigma_{\ExtrinsicState}$ can be considered a normalized timescale of $\ExtrinsicState(t)$ (see section S.4 in the SI Appendix). The computation of the effective noise and its dependency on the environmental timescale is illustrated in Fig.~3.

\begin{figure}
	\centering
	\includegraphics[width=0.9\columnwidth]{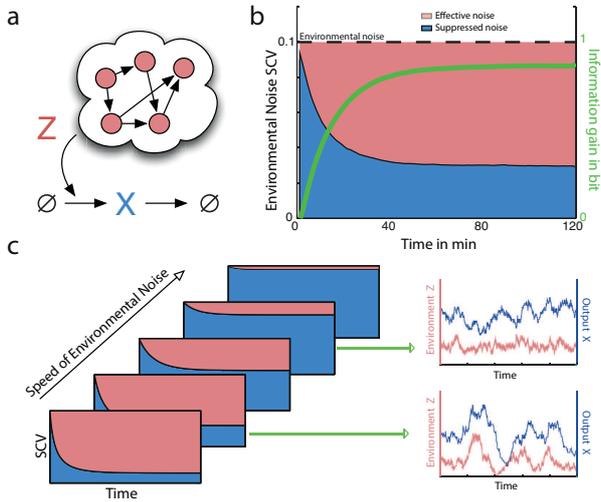}
	\caption{Propagation and suppression of environmental fluctuations. (a) Linear birth-death process in a fluctuation environment. The birth-rate is assumed to be linearly modulated by an environmental stochastic process $\ExtrinsicState$. (b) Calculation of suppressed and effective noise. Individual components were computed analytically by solving an ordinary differential equation (see main text). For orientation, we also show the information gain between $\ExtrinsicState(t)$ and $\ExtrinsicState(t) \mid \StateInt_{t}$, computed using the marginal simulation algorithm (green); it can be understood as the gain in information about $Z$ through observing $X$ and it exhibits a monotone relationship with the effective noise. (c) Relation between the effective noise and the speed of the environmental fluctuations. Noise contributions were computed by numerically solving the ODE from Eq.\ref{eq:ExpectedMoments} for different values of $\theta$ (i.e., timescales).}
		\label{fig:VarianceDecomposition}
\end{figure}

\subsection*{The slow noise approximation (SNA)}
The effective noise can be understood as a measure of how strong $\ExtrinsicState$ impacts $\IntrinsicState$. Only in the special case of a very slow or constant environment, i.e., $\DifferentialOperator_{i}  \approx  0$, we see from Eq.\ref{eq:ExpectedMoments} that for large $t$, $\Var{\Expect{\ExtrinsicState(t) \mid \StateInt_{t}}}\rightarrow \Var{\ExtrinsicState(t)}$, i.e., all variability in $\ExtrinsicState$ is transferred to $\IntrinsicState$. Hence, a more noisy but fluctuating environment may induce a similar (or even the same) effective noise in $\IntrinsicState$ than a random but fixed environment of the same variance. Consequently, when looking at only snapshot data for $\IntrinsicState$ one can generally not infer whether the environment is constant or fluctuating. On the other hand, this implies that we may well approximate the impact of a complicated and dynamically changing environment by a simple random variable of appropriate variance. More specifically, we demand for an equivalent constant environment $\bar{\ExtrinsicState}$ such that $\Var{\bar{\ExtrinsicState}} \equiv \Var{\Expect{\ExtrinsicState(t) \mid \StateInt_{t}}}$,
where $\Var{\Expect{\ExtrinsicState(t) \mid \StateInt_{t}}}=\sigma^{2}$ is the effective noise of the original, fluctuating environment $\ExtrinsicState$ at stationarity. Let us again consider the birth-death process of Fig.~3a and set the birth rate to one such that any scaling is subsumed in the environmental process $Z$. With $\IntrinsicState_{0}=0$, the abundance of the birth death process at any time is given by $\IntrinsicState(t) = R_{b}(t) - R_{d}(t)$ with $R_{b}(t)$ and $R_{d}(t)$ as counting processes for the birth and death reaction, respectively.
We show in section S.5.1 in the SI Appendix that the marginal birth hazard is approximately given by
\begin{equation}
	\lambda_{b}(\StateInt_{t}) = \lambda_{b}(R_{b}(t), t) \approx \frac{\mu^{2} + \sigma^{2} R_{b}(t)}{\mu + \sigma^{2} t} ,
	\label{eq:ApproximateHazard}
\end{equation}
with $\mu = \Expect{\ExtrinsicState(t)}$ the unconditional mean and $\sigma^{2}$ the effective noise of $\ExtrinsicState$, whereas the expression becomes exact for constant and infinitely fast environments.  Note that the marginal hazard does not depend on the full history, but only the number of birth-reactions $R_{b}(t)$ up to time $t$\footnote{That is, $R_{b}(t)$ is a sufficient statistic for evaluating the conditional expectation $\Expect{\ExtrinsicState(t) \mid \StateInt_{t}}$. }. In relation to QSS, which assumes that no fluctuations of $Z$ are propagated to $X$, the found equivalent constant environment with the proper effective noise provides a better approximation for a decoupled simulation of environment and process of interest than QSS.    

Using the effective noise, we now aim to find a master equation, which describes the time-evolution of the marginal probability distribution $P(\intrinsicState, t)$. Since $\lambda_{b}$ depends on $R_{b}(t)$ rather than $\IntrinsicState(t)$, it appears natural to formulate the master equation in $R_{b}(t)$ and $R_{d}(t)$ as well. We remark that since the uncoupled dynamics are non-Markovian, they do not satisfy a conventional master equation. Instead, such processes are described by \textit{generalized master equations} (GME) that can account for memory effects in the dynamics (see S.5 in the SI Appendix for a general derivation and discussion). For the example considered here, one can show that the probability distribution $P(r_{b}, r_{d}, t)$ satisfies a GME of the form
\begin{equation}
	\begin{split}
		&\frac{\mathrm{d}}{\mathrm{d}t}P(r_{b},  r_{d}, t) =  \frac{\mu^{2} + \sigma^{2} (r_{b} - 1)}{\mu + \sigma^{2} t} P(r_{b}-1, r_{d}, t)  \\
		&\quad +  \kparameter_{d}  \left[r_{b} - r_{d} + 1\right] P(r_{b}, r_{d} - 1, t) \\
		&\quad -  \left(\frac{\mu^{2} + \sigma^{2} r_{b}}{\mu + \sigma^{2} t} +  \kparameter_{d}  \left[ r_{b} - r_{d} \right] \right)P(r_{b}, r_{d}, t),
		\label{eq:MESNA}
	\end{split}
\end{equation}
that can be solved analytically using generating functions (see S.5.1 in the SI Appendix). From $P(r_{b}, r_{d}, t)$ we compute the distribution of $\IntrinsicState$ as 
\begin{equation}
	\begin{split}
	P(\intrinsicState, t) &= \sum_{r_{b} =\intrinsicState}^{\infty} P(r_{b}, r_{b} - \intrinsicState, t) \\
	&=\mathcal{NB}\left(x; \frac{\mu^{2}}{\sigma^{2}}, \frac{ \kparameter_{d} \mu e^{\kparameter_{d} t}}{ \kparameter_{d} \mu e^{\kparameter_{d}t}+\left(e^{ \kparameter_{d} t} - 1\right) \sigma^{2}}  \right),
	\label{eq:NegBinomial}
	\end{split}
\end{equation}
i.e., a negative binomial distribution. Eq. \ref{eq:NegBinomial} provides a surprisingly simple approximate solution for the transient probability distribution of birth death processes in a fluctuating environment. In order to check its validity, we compared the analytical approximate distributions to the ones obtained through SSA for a gene expression model, where the environmental fluctuations are assumed to be due to the mRNA dynamics (see Fig.~4). More specifically, we computed the Kolmogorov distance between the resulting protein distributions as a function the environmental timescale. Apart from the exact correspondence for the limiting time-scales, Fig.~4 indicates that the SNA provides a good approximation regardless of the environmental timescale. 


\begin{figure}
	\centering
	\includegraphics[width=0.9\columnwidth]{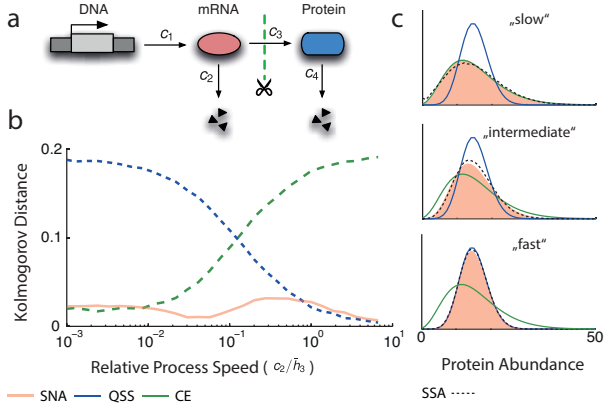}
	\caption{Analytical protein distributions through the slow noise approximation. (a) Two-stage gene expression model. Transcription and translation  are modeled through mass-action kinetics with reaction rate constants $\kparameters_{1}$-$\kparameters_{4}$. Fluctuations on the mRNA are considered environmental and hence, integrated out in order to obtain a one-dimensional stochastic process describing only the protein. (b) Accuracy of the slow noise approximation. The SNA was compared to the QSS- and CE-approximations by means of the Kolmogorov distance between the respective approximate and exact distribution (SSA) as a function of the relative speed of the mRNA fluctuations, i.e., $\kparameters_{2} / \bar{h}_{3}$ and $\bar{h}_{3} = \kparameter_{3} \kparameter_{1} / \kparameter_{2}$. QSS- and CE approximations break down for slow or fast environmental fluctuations respectively, whereas the SNA yields accurate distributions regardless of the mRNA's timescale. (c) Exemplary distributions obtained through the different approaches in three different regimes (slow, intermediate, fast).}
		\label{fig:SNA}
\end{figure}

\section*{Discussion}

There is increasing evidence that models of biochemical networks need to account for both intrinsic and extrinsic noise caused by variations in the intracellular environment. In recent studies, this is done by extending a model's state space by certain environmental species, whose dynamics are described along with the actual system of interest. In particular, the resulting system dynamics are described and studied \textit{conditional} on a particular history of the environment and thus, do not provide a coherent description of a dynamical system subject to extrinsic noise. In this work, we derived and analyzed a novel process framework, which is able to describe just the system of interest as if it was still embedded into its environment. In that sense, it permits a mathematically exact way to analyze small parts of networks in an \textit{uncoupled} fashion. 

Several recent studies rely on the extreme assumptions that the environmental fluctuations are either infinitely fast or slow. While both strategies may in fact lead to strongly simplified and tractable models, they are characterized by significant approximation errors when considering intermediate environmental timescales (see e.g., Fig.~4b). The approach proposed here allows to uncouple a reaction network from its surrounding environment regardless of the latter's timescale. 
In that sense, the approach is fully general although practical implementations may rely on efficient but approximate solutions of the discussed filtering problem.

In the context of Monte Carlo simulation the decoupled process can yield a significant reduction in computational effort when compared to standard SSA -- especially if the environmental network is costly to simulate. This highlights the role of the provided framework as a general tool to split stochastic biochemical networks into individual parts that are easier to simulate. We believe that it will aid in turning stochastic modeling and simulation techniques more large-scale and more faithful to \textit{in vivo} conditions, where significant environmental fluctuations are present. Moreover, the framework can be used in the model-based design of novel circuit motifs in synthetic biology and is related to the notion of retroactivity \cite{Jayanthi2013}.  

We further demonstrate that the uncoupled dynamics provide a novel analytical tool to study how environmental stochasticity is propagated along coupled reaction networks. For instance, we have shown that the total environmental noise splits up into two terms: one corresponding to the noise that is suppressed and a second term that quantifies the effective noise that is \textit{sensed} by the target network. 

In \cite{Lestas2010} the authors derive a lower bound on a network's ability to suppress fluctuations and show its immediate relation to the uncertainty at which those fluctuations can be estimated -- similar to what we defined as effective noise. The methods proposed here allow to not only bound, but fully determine both the suppressed and effective environmental noise. 
Our results further indicate that two environments with very different timescales may impact a network in a similar way. For instance a fixed but random environment may yield the same effective noise as a fluctuating environment with larger variance. 
%
Along those lines, we derived a simple but widely applicable approximation of the transient probability distribution for birth death processes subject to environmental noise. It is based on the idea to approximate a fluctuating environmental process by a simple random variable that impacts the birth death process in an equivalent way. 
In order to solve for the transient probability distribution we derived a novel generalized master equation for this non-Markovian process. 

\section*{Methods}
\everymath{\footnotesize\scriptstyle}


\subsection*{Normalized filtering distribution.} 
The unnormalized filtering distribution from Eq.\ref{eq:PosteriorProbability} does not sum up or integrate to one and therefore, cannot be used to derive statistics such as moments and so forth. We show in section S.2 in the SI Appendix that the normalized filtering distribution is given by
\begin{equation}
	\begin{split}
	\mathrm{d}\pi(z, t) &= \Big( \KolmogorovOperator \pi(z, t)  - \kparameter_{k} \left[z - \momentOrder{1}(t)  \right] \pi(z, t)  \Big)\mathrm{d}t \\
	&\quad+\left[\frac{z - \momentOrder{1}(t)}{\momentOrder{1}(t)}\right] \pi(z, t) \mathrm{d}R_{k}(t),
	\end{split}
	\label{eq:NormalizedPosteriorProbability}
\end{equation}
which -- due to a dependency on the mean $\momentOrder{1}(t)$ -- is difficult to integrate numerically. On the other hand, moment dynamics are straight-forward to derive such as described in the following section.

\subsection*{Conditional moment dynamics.}
The $i$-th order non-central moment is computed by multiplying both sides of Eq.\ref{eq:NormalizedPosteriorProbability} with $\extrinsicState^{i}$ and summing (or integrating) over all $\extrinsicState \in \ExtrinsicSpace$, i.e.,
\begin{equation}
	\begin{split}
	 \sum_{z\in \ExtrinsicSpace}  z^{i} \mathrm{d}  \pi(z, t) &= \sum_{z\in \ExtrinsicSpace}  z^{i} \Big( \KolmogorovOperator \pi(z, t) - \kparameter_{k} \left[z - \momentOrder{1}(t) \right] \pi(z, t) \Big) \mathrm{d}t \\
	&+ \sum_{z\in \ExtrinsicSpace}  z^{i} \left[\frac{z - \momentOrder{1}(t)}{\momentOrder{1}(t)}\right] \pi(z, t) \mathrm{d}R_{k}(t) \\
	&= \left[\DifferentialOperator_{i}(t) - \kparameter_{k} (\momentOrder{i+1}(t) - \momentOrder{1}(t) \momentOrder{i}(t)) \right]\mathrm{d}t \\
		&\quad\quad\quad\quad + \frac{\momentOrder{i+1}(t) - \momentOrder{1}(t)\momentOrder{i}(t)}{\momentOrder{1}(t)}\mathrm{d}R_{k}(t),
	\end{split}
	\label{eq:MomentDerivation}
\end{equation}
with $\DifferentialOperator_{i}(t) = \sum_{z\in\ExtrinsicSpace} z^{i} \KolmogorovOperator \pi(z, t)$. The computation of moments in case of multivariate environments is obtained analogously.

\subsection*{Marginal simulation algorithm.} As indicated in the main text, the uncoupled dynamics can be simulated using any stochastic simulation algorithm that can deal with time-varying hazard functions. Although more efficient variants might be possible, we make use of a first-reaction method \cite{Anderson2007} that we combine with a thinning algorithm \cite{Lewis1979}. The first reaction method is based on drawing waiting times $\tau_{1}, \ldots, \tau_{N}$ for each of the $N$ reaction channels and then picking the reaction corresponding to the minimum of those waiting times. First we remark that only the reactions involving $\ExtrinsicState(t)$ will be affected by the decoupling scheme and hence, all other reactions will retain their exponential waiting time distributions. In order to simulate from the non-exponential waiting time distribution from Eq.\ref{eq:WaitingTimeGeneral}, we apply a thinning algorithm given by the following steps:
\begin{enumerate}
\footnotesize
	\item Set $\tau = t$.
	\item Simulate $\hat{\tau} \sim \mathrm{Exp}(\hat{\lambda})$.
	\item Set $\tau = \tau + \hat{\tau}$.
	\item Draw $u\sim \mathcal{U}(0, 1)$. If $u\leq \frac{\momentOrder{1}(\tau) g_{k}(\stateInt(\tau))}{\hat{\lambda}}$ return $\tau$. Else, go back to step 2.
\end{enumerate}
Note that the tuning parameter $\hat{\lambda}$ has to be chosen such that $\hat{\lambda} \geq \momentOrder{1}(\tau) g_{k}(\stateInt(\tau))$ for all $\tau \in [0, T]$, where $T$ is the simulation interval.

\subsection*{Acknowledgements.}
C.Z. and H.K. acknowledge the support from the Swiss National Science Foundation, grant no. PP00P2 128503 and SystemsX.ch, respectively.


{\footnotesize
	\begin{thebibliography}{10}

\bibitem{Mcadams1997}
McAdams HH, Arkin A
\newblock (1997) Stochastic mechanisms in gene expression.
\newblock \emph{Proc Natl Acad Sci USA} 94:814--819.

\bibitem{Elowitz2002}
Elowitz MB, Levine AJ, Siggia ED, Swain PS
\newblock (2002) {Stochastic gene expression in a single cell.}
\newblock \emph{Science} 297:1183--6.

\bibitem{Colman-Lerner2005}
Colman-Lerner A, {et~al.}
\newblock (2005) {Regulated cell-to-cell variation in a cell-fate decision
  system.}
\newblock \emph{Nature} 437:699--706.

\bibitem{oshea}
Raser JM, O'Shea EK
\newblock (2004) Control of stochasticity in eukaryotic gene expression.
\newblock \emph{Science} 304:1811--1814.

\bibitem{Hilfinger2011}
Hilfinger A, Paulsson J
\newblock (2011) {Separating intrinsic from extrinsic fluctuations in dynamic
  biological systems.}
\newblock \emph{Proc Natl Acad Sci USA} 108:12167--12172.

\bibitem{Swain2002}
Swain PS, Elowitz MB, Siggia ED
\newblock (2002) Intrinsic and extrinsic contributions to stochasticity in gene
  expression.
\newblock \emph{Proceedings of the National Academy of Sciences}
  99:12795--12800.

\bibitem{Ruess2013}
Ruess J, Milias-Argeitis A, Lygeros J
\newblock (2013) Designing experiments to understand the variability in
  biochemical reaction networks.
\newblock \emph{Journal of The Royal Society Interface} 10.

\bibitem{Zechner2012}
Zechner C, {et~al.}
\newblock (2012) Moment-based inference predicts bimodality in transient gene
  expression.
\newblock \emph{Proc Natl Acad Sci USA} 109:8340--8345.

\bibitem{Zechner2014}
Zechner C, Unger M, Pelet S, M. P, Koeppl H
\newblock (2014) Scalable inference of heterogeneous reaction kinetics from
  pooled single-cell recordings.
\newblock \emph{Nat Methods} 11:197--202.

\bibitem{Finkenstadt2013}
Finkenst{\"a}dt B, {et~al.}
\newblock (2013) Quantifying intrinsic and extrinsic noise in gene
  transcription using the linear noise approximation: An application to single
  cell data.
\newblock \emph{The Annals of Applied Statistics} 7:1960--1982.

\bibitem{Hasenauer2011}
Hasenauer J, {et~al.}
\newblock (2011) Identification of models of heterogeneous cell populations
  from population snapshot data.
\newblock \emph{BMC Bioinformatics} 12:125.

\bibitem{Shahrezaei2008}
Shahrezaei V, Ollivier JF, Swain PS
\newblock (2008) {Colored extrinsic fluctuations and stochastic gene
  expression.}
\newblock \emph{Mol Syst Biol} 4:196.

\bibitem{Johnson1977}
N.~Johnson SK
\newblock (1977) \emph{Urn Models and Their Application}
\newblock (Wiley \& Sons, New York).

\bibitem{Koeppl2012}
Koeppl H, Zechner C, Ganguly A, Pelet S, Peter M
\newblock (2012) Accounting for extrinsic variability in the estimation of
  stochastic rate constants.
\newblock \emph{Int J Robust Nonlin} 22:1103--1119.

\bibitem{Arkin2003}
Rao CV, Arkin AP
\newblock (2003) Stochastic chemical kinetics and the quasi-steady-state
  assumption: Application to the gillespie algorithm.
\newblock \emph{The Journal of Chemical Physics} 118:4999--5010.

\bibitem{AalenBook}
Aalen OO, Borgan, Gjessing HK
\newblock (2008) \emph{Survival and event history analysis: a process point of
  view}
\newblock (Springer Verlag).

\bibitem{crisanbook}
Bain A, Crisan D
\newblock (2009) \emph{Fundamentals of stochastic filtering}
\newblock (Springer, New York).

\bibitem{Kurtz2011}
Anderson DF, Kurtz TG
\newblock (2011) in \emph{Design and Analysis of Biomolecular Circuits}
\newblock (Springer), pp 3--42.

\bibitem{Snyder}
Snyder DL, Miller MI
\newblock (1975) \emph{Random Point Processes in Time and Space}
\newblock (Wiley \& Sons, New York).

\bibitem{Elliott2005}
Elliott RJ, Malcolm WP
\newblock (2005) General smoothing formulas for markov-modulated poisson
  observations.
\newblock \emph{Automatic Control, IEEE Transactions on} 50:1123--1134.

\bibitem{Anderson2007}
Anderson DF
\newblock (2007) {A modified next reaction method for simulating chemical
  systems with time dependent propensities and delays.}
\newblock \emph{J Chem Phys} 127:214107.

\bibitem{Lewis1979}
Lewis PAW, Shedler GS
\newblock (1979) Simulation of nonhomogeneous {P}oisson processes by thinning.
\newblock \emph{Naval Research Logistics Quarterly} 26:403--413.

\bibitem{Rosenfeld2005}
Rosenfeld N, Young JW, Alon U, Swain PS, Elowitz MB
\newblock (2005) Gene regulation at the single-cell level.
\newblock \emph{Science} 307:1962--1965.

\bibitem{Volfson2006}
Volfson D, {et~al.}
\newblock (2006) Origins of extrinsic variability in eukaryotic gene
  expression.
\newblock \emph{Nature} 439:861--864.

\bibitem{Zechner2013}
Zechner C, Deb S, Koeppl H
\newblock (2013) Marginal dynamics of stochastic biochemical networks in random
  environments.
\newblock \emph{2013 European Control Conference (ECC)} pp 4269--4274.

\bibitem{Bowsher2013}
Bowsher CG, Voliotis M, Swain PS
\newblock (2013) The fidelity of dynamic signaling by noisy biomolecular
  networks.
\newblock \emph{Plos Comp Biol} 9:e1002965.

\bibitem{Jayanthi2013}
Jayanthi S, Nilgiriwala KS, Del~Vecchio D
\newblock (2013) Retroactivity controls the temporal dynamics of gene
  transcription.
\newblock \emph{ACS Synthetic Biology} 2:431--441.

\bibitem{Lestas2010}
Lestas I, Vinnicombe G, Paulsson J
\newblock (2010) Fundamental limits on the suppression of molecular
  fluctuations.
\newblock \emph{Nature} 467:174--178.

\end{thebibliography}

}

\end{document}